# Do Prescribed Prompts Prime Sensemaking During Group Problem Solving?

Mathew "Sandy" Martinuk[a] and Joss Ives[b]

[a]Department of Physics and Astronomy, University of British Columbia, Vancouver, BC, V6T 1Z1
[b]Department of Physics, University of the Fraser Valley, Abbotsford, BC, V2S 7M8

**Abstract.** Many researchers and textbooks have promoted the use of rigid prescribed strategies for encouraging development of expert-like problem-solving behavior in novice students. The University of British Columbia's introductory algebra-based course for non-physics majors uses Context-Rich problems with a prescribed six-step strategy. We have coded audio recordings of group problem-solving sessions to analyze students' epistemological framing based on the implicit goal of their discussions. By treating the goal of "understanding the physics of the situation" as sensemaking, we argue that prescribed problem-solving prompts are not sufficient to induce subsequent sensemaking discussion.

## INTRODUCTION

Research on expert-novice differences in problem solving have revealed significant differences in how these two groups categorize problems, plan solutions, and check their work. In order to enable students to develop expert-like problem-solving skills, many researchers and textbooks advocate the use of rigidly prescribed problem-solving strategies in introductory physics. [1-5]

One concern about the use of such strategies is that students may treat the problem-solving steps as a list of instructions to follow rather than as individual elements that contribute to overall understanding and a coherent problem solution.

This paper examines how students respond to prescribed problem-solving prompts with a primary research question of: *Are prescribed problem-solving prompts effective at promoting student sensemaking?* In this paper we operationalize the term "sensemaking" as students being observably engaged in conceptual discussion. We analyze audio recordings of small group problem-solving sessions to examine the relationship between prescribed problem-solving prompts and conceptual discussion.

## BACKGROUND

Physics 100 at the University of British Columbia is a 13-week algebra-based introductory physics course offered to science students who require physics credit to complete their degree but who did not take senior high school physics. It is a large lecture-based course which divides approximately 800 students into three sections, each taught by a different instructor. The lecture is supported by weekly labs and mandatory small-group problem-solving sessions. Students in Physics 100 are typically enrolled in science, human kinetics, forestry, food and nutrition science, or various arts programs, with fewer than 1% intending to major in physics.

The course was substantially revised in 2007 to present introductory physics in a real-world context wherever possible. The instructors wanted to demonstrate the relevance of physics to the real world and enable students to solve novel real-world problems. For a more thorough description of the course changes see Martinuk et al. [6]

### Structured Problem Solving

To support the development of students' problem-solving skills a structured problem-solving method was introduced in the Physics 100 lectures and small-group problem-solving sessions. Constraining novices to use more expert-like solution strategies is a widely used practice intended to foster their development of expert problem-solving skills. Among the advocates of explicitly structured problem-solving methods are Heller et al. [1] who use structured worksheets for context-rich questions, Teodorescu's ACCESS protocol [2], van Heuvelen's Active Learning Problem Sheets [3] and many popular introductory Physics textbooks. [e.g. 4,5]

In UBC's Physics 100, a six step problem-solving strategy (see Table 1) was used during lecture and on the small-group problem-solving worksheets. The worksheets presented the problem and then each step followed by a brief description of that step. Steps 2 through 6 were mandatory and each step was graded. In order to familiarize the students with the elements of this strategy, the first five problem-solving sessions were explicit hour-long workshops on each piece of the strategy, similar to the structured approach advocated by Teodorescu. [2]

We expected that having the descriptive prompts for steps 1, 2, 3, and 6 written on their worksheets would encourage students to look beyond their calculations and engage in sensemaking of the problem under consideration.

**TABLE 1.** The prescribed problem-solving steps provided on the small-group problem-solving worksheets.

| Step # | Step Name |
|---|---|
| 1 | Interpret the Problem |
| 2 | Identify Relevant Physics |
| 3 | Model: Identify Assumptions and Relationships |
| 4 | Model: Construct a Diagram |
| 5 | Solve the Problem |
| 6 | Error-checking and Sensemaking |

## METHODOLOGY

To examine students' sensemaking during problem solving we used the idea of Epistemological Framing, which is a person's implicit expectations about knowledge and learning in their current activity and context. An epistemological frame comes with a set of expectations about what kind of knowledge is relevant, what the goal of the activity is, and how progress will be made, and is usually shared by all members of a collaborative group. [7, 8]

We used a modified version of a coding scheme developed by Scherr and Hammer [8] to examine how students were approaching these problems. We refined the scheme in two ways. First we demonstrated that coding using only audio recordings can yield the same frames as using the full video + audio more than 80% of the time. [9] Second we refined Scherr and Hammer's Discussion frame into two categories, which were seen to arise naturally from the implicit goal of students' speech: Procedural Discussion and Conceptual Discussion. Our scheme is summarized on Table 2.

Procedural Discussion has the implicit goal of "figuring out what to do" in order to make progress on the worksheet. Conceptual Discussion has the implicit goal of "figuring out what is happening" in the physics or interpreting the meaning of the physics situation. The Conceptual Discussion frame is where we saw students striving to express, understand, and synthesize new ideas. For this reason, we chose to operationalize the more general term "sensemaking" specifically as "engaging in Conceptual Discussion".

**TABLE 2.** Summary of the Epistemological Framing Coding Scheme.

| Frame | | Description |
|---|---|---|
| Conceptual Discussion | CD | Engaged discussion to understand meaning of physics |
| Procedural Discussion | PD | Engaged discussion to figure out how to proceed or what the professor expects |
| Worksheet Focus | W | Focus on writing on worksheet or directing others' writing |
| TA Focus | TA | Focus on interacting with Teaching Assistant |
| Other / Off-topic | O | Meta-comments, group role negotiations, off-topic discussion |

Using this scheme, the authors achieved greater than 80% inter-rater reliability (IRR) with 100% agreement after discussion. Because the judgments of Conceptual Discussion framing are so central to our analysis, we also conducted an independent test of inter-rater reliability on just this frame. We found 71% IRR before discussion, which was largely due to gradual transitions between the two Discussion frames. This seems to be a common characteristic of these frames but has not yet been incorporated into the coding scheme. For this paper, the authors discussed all instances of Conceptual Discussion and achieved 100% consensus on the Conceptual Discussion codes.

Six episodes of group problem solving were analyzed for this paper. Each episode is a different group working with the same prescribed prompts on one of two problems which had identical mathematical structure with slightly different real-world cover stories. These problems were chosen because they are typical examples of the context-rich problems used in this course. Each episode was transcribed and coded for students' epistemological framing throughout. The time when students reached each problem-solving prompt was also recorded. In instances where it was unclear when the group had reached a particular prompt, they were excluded from analysis of that prompt.

In order to examine the relationship between structured prompts and sensemaking we examine the students' framing after they encounter a given worksheet prompt. Because of the complexity in ascribing direct causality between a prompt and subsequent conceptual discussion, we made the generous assumption that any Conceptual Discussion occurring during the segment after one prompt and before the next was the result of that prompt.

**TABLE 3.** The percentage of time spent in the Conceptual Discussion frame during different problem-solving segments. "Overall Average" indicates the total time spent in the Conceptual Discussion frame over the total time spent in that segment. "NA" indicates that it was not possible to tell when the group reached that prompt.

|  | 1. Interpret | 2. Relevant Physics | 3. Assumptions | 4. Diagram | 5. Solve | 6. Error-Checking | Overall Average |
|---|---|---|---|---|---|---|---|
| Group #1 | 41% | 0% | 10% | 0% | 0% | 0% | 4% |
| Group #2 | NA | 7% | 6% | 0% | 0% | 0% | 2% |
| Group #3 | 13% | 0% | 18% | NA | 0% | 4% | 9% |
| Group #4 | 10% | 0% | 14% | 0% | 38% | 7% | 14% |
| Group #5 | 0% | 0% | 10% | 4% | 5% | 0% | 4% |
| Group #6 | 0% | 2% | 19% | 6% | 10% | 24% | 9% |
| Overall Average | 11% | 2% | 13% | 4% | 8% | 5% |  |

## RESULTS AND DISCUSSION

To investigate the effectiveness of each prompt we examined the number of groups that engaged in Conceptual Discussion during each segment and the percentage of time they spent engaged in Conceptual Discussion. These results are summarized in Table 3.

There is a high variability in results, both between groups and between prompts. This is unsurprising considering the variation in students' knowledge, beliefs, group dynamics, and the complexity of ways that people may respond to a printed prompt.

### Conceptual Discussion Benchmark

In order to put these Conceptual Discussion percentages in context we needed to estimate what a reasonable upper limit of Conceptual Discussion for a "good" group interaction might be. To develop a benchmark we identified a sub-section of one episode where a well-functioning group (#4) went through a sequence of raising a new question, discussing it, and recording their result. This section was chosen specifically because we felt that it was the best example of engaged conceptual discussion available in our dataset. The frames in this section are illustrated in Figure 1.

The sequence used follows the Solve prompt, where the group has just realized that dynamics are needed for solving the problem. During the first three minutes, the group creates and records a new model for their system. During minutes four to six they define a coordinate system and re-interpret the problem's given information for their new model. During minutes six through eight the group conducts their calculations, and then they interpret their answer and record their solution.

Based on this episode, we estimate that a well functioning group could spend as much as 38% of their time on Conceptual Discussion while working on a quantitative context-rich problem.

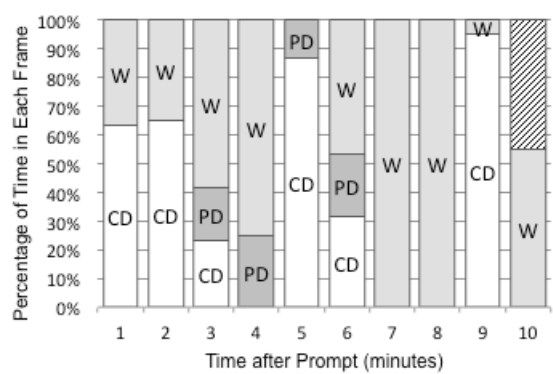

**FIGURE 1.** The proportion of frames for the Conceptual Discussion benchmark example. The overall frame ratios are Worksheet (W) = 53%; Procedural Discussion (PD) = 8%; and Conceptual Discussion (CD) = 38%. The cross-hatched area indicates the end of the benchmark section during the tenth minute.

### Assumptions Segment Has Most Conceptual Discussion

The Assumptions segment was the only one where all six student groups engaged in Conceptual Discussion. The Interpret segment shows a similar percentage of Conceptual Discussion, but that is artificially inflated by one group who spends a long period of time discussing the meaning of a formula they have just copied out of their textbook because of a superficial similarity to the current problem. By contrast, most of the Conceptual Discussion in the Assumptions segment corresponds to students debating the meaning and merit of various assumptions; activities which we believe represent authentic sensemaking.

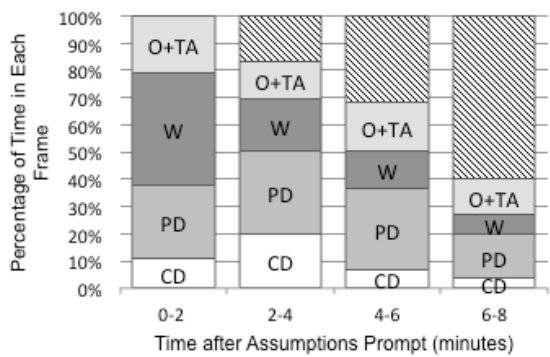

**FIGURE 2.** The proportion of frames after the "Identify Assumptions and Relationships" prompt, totaled for all six student groups. The cross-hatched area corresponds to groups that have completed this problem-solving segment.

*Very Little Conceptual Discussion Overall*

Treating the above benchmark as an estimate of the upper limit on the percentage of time spent in the Conceptual Discussion frame during quantitative group problem solving we see that the majority of segments fall well short of this upper limit. This suggests that these prompts are not, in general, effective in promoting sensemaking.

In particular, prompts 1, 2, and 6 were intended to promote sensemaking, but at best half of the student groups engage in Conceptual Discussion during these segments and the average proportion of time spent in Conceptual Discussion was only 6%, which is very low compared to the upper limit estimate of 38%.

The low rate of Conceptual Discussion during the Error-Checking and Sensemaking segment despite explicitly teaching sensemaking strategies was especially surprising. Investigation of the students' discourse showed that some groups engaged in conceptual sensemaking immediately following their calculation instead of when prompted by the worksheet, while other groups treated the prompt as an external requirement to be satisfied and engaged in only a cursory examination of their result.

## DISCUSSION

Despite looking at the entire segment following each problem-solving prompt, we find that few students engage in Conceptual Discussion after a prompt and those that do spend very little of time on it. While it is possible that this is due to the influence of the questions studied, we have no evidence that these questions are in any way atypical of context-rich physics problems. This suggests that prescribed problem-solving strategies are ineffective at prompting high levels of sensemaking for these problems.

While Dufresne et. al. [10] demonstrated improved problem-solving performance by constraining novices to engage in expert-like problem-solving behavior, we contend that their use of a step-by-step computer guide does not reflect common classroom conditions. We speculate that in a classroom context students perceive structured problem-solving prompts as a list of conditions to be satisfied for marks which keeps their focus on what is required to earn marks rather than on making sense of their process and their answers. In this speculation we agree with Heckler, [11] who saw decreased performance as a result of explicitly prompting students to produce a free-body diagram prior to solving a dynamics problem and suggests that this prompt "cued some students to the mindset that constructing the diagram and solving the problem are two separate tasks."

Our most successful prompt was the explicit requirement to state modeling assumptions, which prompted Conceptual Discussion from every group studied, and had the highest percentage of Conceptual Discussion overall. We believe that this prompt is successful only because reconciliation between formal physics and everyday intuition is required to complete this task, and so a certain amount of Conceptual Discussion is necessary.

## ACKNOWLEDGMENTS

This work was supported by the Carl Wieman Science Education Initiative and the UBC Department of Physics and Astronomy. We would also like to thank Georg Rieger, Andrzej Kotlicki, and Fei Zhou.

## REFERENCES

1. P. Heller, R. Keith, and S. Anderson, *Am. J. Phys* **60**, 627-636 (1992).
2. R. Teodorescu, Ph.D. dissertation, The George Washington University, 2009.
3. A. van Heuvelen, *Am. J. Phys* **59**, 898-907 (1991).
4. R. Knight, *Physics for Scientists and Engineers: A Strategic Approach* (Addison Wesley, Boston, 2008)
5. H. Young, R. Freedman, and L. Ford, *University Physics with Modern Physics* (Addison Wesley, Boston, 2008).
6. M. Martinuk, R. Moll, and A. Kotlicki, *Phys. Teach.* **48**, 413–415 (2010).
7. D. Tannen, Framing in Discourse (Oxford University Press, New York, 1993).
8. R. Scherr and D. Hammer, *Cognition and Instruction*, **27**, 147-174 (2009).
9. M. Martinuk and R. Scherr (unpublished)
10. R. Dufresne, W. Gerace, and P. Hardiman, *The Journal of the Learning Sciences* **2**, 307–331 (1992).
11. A. Heckler, *International Journal of Science Education* **32**, 1829–1851 (2010).